\def\arxiv{}

\ifdefined\thesis
    \documentclass[reprint,NumberedRefs]{JASA_blank}
    \def\showfigures{}
\else
\ifdefined\reprint
    \documentclass[reprint,NumberedRefs]{JASA}
    \def\showfigures{}
\else
\ifdefined\reviewers
    \documentclass[preprint,NumberedRefs,TurnOnLineNumbers,trackchanges]{JASA}
    \def\showfigures{}
\else
\ifdefined\arxiv
    \documentclass[twocolumn,amsmath,amssymb,reprint,aip,floatfix]{revtex4-1}
    \def\showfigures{}
    \usepackage{graphicx}
    \usepackage{bm}
    \usepackage{amsfonts}
    \usepackage{latexsym}%
    \usepackage{xcolor}
    \usepackage[bookmarks=false,         
         pdfnewwindow=true,      
         colorlinks=true,    
    final=true
     ]{hyperref}
    \newdimen\reprintcolumnwidth
    \global\reprintcolumnwidth=246pt
    \def\dodoi#1{doi: \href{https://doi.org/#1}{\nolinkurl{#1}}}
    \def\dourl#1{\href{http://#1}{\nolinkurl{#1}}}
\else 
    \documentclass[preprint,NumberedRefs]{JASA}
    \def\appendcaptions{}
\fi
\fi
\fi
\fi

\graphicspath{{./fig/}}

\newcommand{\T}{^{\sf T}}
\renewcommand{\H}{^{\sf H}}
\newcommand{\kz}{^{k}} 
\newcommand{\kn}{^{k+1}}
\newcommand{\mb}{\mathbf}
\newcommand{\obs}{{\mathrm{obs}}}
\newcommand{\ldconv}{\circledast\overset{L}{\cdots}\circledast} 
\renewcommand{\th}{^\text{th}} 
\renewcommand{\labelenumi}{\theenumi}
\usepackage{letltxmacro}
\LetLtxMacro{\originaleqref}{\eqref}
\renewcommand{\eqref}[1]{Eq.~(\ref{#1})}

\begin{document}

\title[Sound field reconstruction using convolutional waves]{A convolutional plane wave model for sound field reconstruction}

\author{Manuel Hahmann}
\author{Efren Fernandez-Grande}

\email{efgr@dtu.dk}
\affiliation{Acoustic Technology Group, Department of Electrical Engineering, Technical University of Denmark, Building 352, \O rsteds Plads, 2800 Kgs.~Lyngby, Denmark}

\begin{abstract} 
    Spatial sound field interpolation relies on suitable models to both conform to available measurements and predict the sound field in the domain of interest.
    A suitable model can be difficult to determine when the spatial domain of interest is large compared to the wavelength or when spherical and planar wavefronts are present or the sound field is complex, as in the near-field.
    To span such complex sound fields, the global reconstruction task can be partitioned into local subdomain problems.
    Previous studies have shown that partitioning approaches rely on sufficient measurements within each domain, due to the higher number of model coefficients.
    This study proposes a joint analysis of all local subdomains, while enforcing self-similarity between neighbouring partitions.
    More specifically, the coefficients of local plane wave representations are sought to have spatially smooth magnitudes.
    A convolutional model of the sound field in terms of plane wave filters is formulated and the inverse reconstruction problem is solved via the alternating direction method of multipliers.
    The experiments on simulated and measured sound fields suggest, that the proposed method both retains the flexibility of local models to conform to complex sound fields and also preserves the global structure to reconstruct from fewer measurements.
\end{abstract}

\maketitle

\hypertarget{introduction}{%
    \section{Introduction}\label{introduction}}


Sound field reconstruction methods enable spatial interpolation of sound fields from a set of discrete measurements.
Such spatial characterization of sound fields is key in applications such as 
sound field analysis, \cite{jacobsen2010a,verburg2018a,haneda1999a,mignot2013a,mignot2014a,nolan2019a,witew2017a,brandao2022a}
sound field control, \cite{heuchel2020a,caviedes2019a,heuchel2018a,betlehem2005a,moller2019a}
in simulation software (interpolation from a coarser to a finer grid),\cite{borrel2021a}
and for navigation of a sound field in auralization and spatial audio.\cite{tylka2017a,tylka2020a,winter2014a,schultz2013data,fernandez2021a}
Often, the sound field at hand is dominated by wavefronts of specific geometry, and a matching propagation model is sufficient to approximate the measurements and interpolate the sound field.
Typical approaches use for example plane waves \cite{verburg2018a,mignot2014a,nolan2019a,jacobsen2013a,moiola2011a} or spherical harmonics \cite{caviedes2019a,betlehem2005a,pezzoli2022sparsity} (in free or near field).


When considering areas significantly larger than the acoustic wavelength, parts of the sound field can show significant influence of reflections, scattering, diffraction or varying wavefront curvature.\cite{witew2017a,heuchel2020a,brandao2022a,fernandez2021a}
To model fields across such large domains, typical approaches include wave expansions with a high number of terms, or dividing the global domain into smaller local subdomains.
For example, plane waves have been proposed to interpolate between two local spherical harmonic decompositions.\cite{tylka2020b}
Another approach is to analyse the sound field in terms of independent overlapping subdomains, as it is common in other disciplines like image processing.\cite{elad2006a}
In acoustic field analysis, subdomain representations using plane waves (also called ray space analysis) have been explored.\cite{markovic2016a,markovic2013a,hahmann2021spatial}
Sparse subdomain representations have been examined for beamforming,\cite{jin2017a} and sound field reconstruction using plane waves\cite{yu2021upscaling} and functions learned from measured sound fields.\cite{hahmann2021spatial} 

Such locally variant representations increase the number of model coefficients to span more complex observations.
However, independent local representations of sound fields ignore the continuous nature of wave propagation.
Even diffuse fields, which exhibit the shortest possible correlation length, are commonly described as a superposition of infinitely many plane waves with random incidence angles and phases.\cite{morse1944a,schroeder1954a,pierce1981a}
It is therefore reasonable to assume spatial similarity between sound field representations in overlapping subdomains.


Convolutional models express a given field in terms of a set of subdomain-size filters, convolved with a spatial coefficient map.
The spatial coefficient map preserves the spatial context of subdomains and allows for joint analysis of neighbouring partitions, thereby capturing both the fine structure and large-scale features of the field.\cite{papyan2017a,wohlberg2018a,bianco2018a}
Such convolutional approaches often exploit sparsity to find optimal local filter coefficients.
They are then known as convolutional or shift-invariant sparse coding in audio and image processing\cite{grosse2007a,m2008a,batenkov2017a} and parallels to convolutional neural networks exist.\cite{papyan2017b}
Convolutional analysis has previously been proposed for beamforming\cite{cohen2018sparse}, also in form of neural networks for spatial sound field interpolation\cite{llu2020a} and source localization.\cite{grumiaux2021a}

In this study, we express a monochromatic sound field in terms of a locally variant planar wave model.
In a convolutional formulation, we enforce continuity between local representations to exploit observations in neighboring subdomains.
In this way, we not only accommodate local phenomena, but also capture the global structure of the sound field (by context between neighbours).
Specifically, we enforce spatially smooth coefficients in a joint analysis of all local plane wave representations.
In addition to global continuity, we also require sparsity within each local subdomain.
Because the spatial frequencies are bandlimited, sparse approximations enable
reconstructions even from few observations within each local
subdomain.\cite{verburg2018a,gerstoft2018a,candes2008a}

To test the proposed continuous convolutional plane wave model, we reconstruct:
in Sec.~\ref{1dmonopole} a simulated sound field in near field of a monopole, radially across a linear array,
in Sec.~\ref{2dmonoplane} a simulated field of a monopole interferring with a plane wave across a 2D aperture and
in Sec.~\ref{classroom} the experimentally captured reverberant high-frequency sound field in a classroom across a large 2D aperture.
The reconstruction is formulated as an Alternating Direction Method of Multipliers (ADMM) problem.\cite{boyd2010admm} 
Where applicable, steps are solved in frequency domain, where the convolution transforms to a multiplication.\cite{heide-2015-fast,wohlberg2016b}
To enable reconstruction across a limited and sparsely sampled aperture, mask decoupling is applied.\cite{wohlberg2016a,wohlberg2017a}
As benchmarks, global and independent local plane wave reconstructions are included in the tests.


\hypertarget{theory}{%
    \section{Theory}\label{theory}}

This section explains the sound field modelling and reconstruction approaches included in this study:
the conventional linear superposition model in \ref{global_rec}, 
the partitioning of the reconstruction domain into independent, overlapping subdomains in \ref{local_mod},
the convolutional model to facilitate spatially smooth local coefficients in \ref{conv_mod}, 
its solution via ADMM in \ref{conv_admm},
and the assessment of reconstructed sound fields in \ref{eval}.

\hypertarget{global_rec}{%
\subsection{Global sound field model and reconstruction}\label{global_rec}}
\noindent

The true acoustic pressure $\mb{p}\in \mathbb{C}^N$ at frequency $f$ and $N$ positions $\mb r \in \mathbb{R}^{3}$ within a domain $\Omega$ is assumed to be modelled as a linear combination of basis functions
\begin{equation}
    \mb{p = H x} \ ,
    \label{eq:linear_model}
\end{equation}

where $\mb x \in \mathbb C^M$ are the coefficients and $\mb H \in \mathbb C^{N \times M}$ contains the $M$ basis functions.
For example, plane propagating waves $e^{-\mathrm{j} \mb k\T \mb r}$ are often used to model reverberant or far-fields, where $\mb k$ is the wavenumber vector ($\|\mb k\|_2 = 2\pi /\lambda$, $\lambda$ is the wavelength).
In the case of plane waves, the $n,m\th$ element of $\mb H$ is $e^{-\mathrm{j} \mb k_m\T \mb r_n}$, with $\mb r_n$ denoting the $n\th$ position and $\mb k_m$ the wavenumber vector from the $m\th$ incidence direction.

For the equation to hold, $\mb H$ must span the observed sound pressure field $\mb p$ over the complete domain $\Omega$.
An observation of the sound pressure field is 
\begin{equation}
    \mb p_\obs = \mb {M p + n} = \mb H_\obs \mb{x + n} \ ,
    \label{eq:observation}
\end{equation}
where $\mb{M}\in \{0,1\}^{N_\obs\times N}$ is a binary mask selecting the $N_\obs$ available observations from the sound field $\mb p$.
$\mb H_\obs = \mb {M H}$ is the model at the observed positions and $\mb n$ is an error vector, that accounts for measurement noise and model error.

An optimal set of coefficients $\hat{\mb x}$ is found by inversion of \eqref{eq:observation}.
To arrive to a stable solution, regularization is necessary as $\mb H$ is typically ill-conditioned or rank-deficient.\cite{hansen1998a} 
Typically, a structure in the coefficients is imposed, such as in
\begin{equation}
    \hat{\mb x} = \text{arg} \min_{\mb x} \; \left\lVert \mb{H}_\obs \mb{x} - \ \mb{p}_\obs \right\rVert_2^2 + \beta \
    \| \mb{x} \|_1
    \label{eq:compressive_sensing} \ ,
\end{equation}
in which case a $\ell_1$-norm penalty is applied to promote a sparse structure in the coefficients and $\hat{\cdot}$ denotes an estimate.
The sound field is then reconstructed as
\begin{equation}
    \hat{\mb p} = \mb H \hat{\mb x}
    \label{eq:global_rec} \ ,
\end{equation} 
where $\hat{\mb p} \in \mathbb{C}^{N}$ is the reconstructed sound pressure at $N$ reconstruction positions.

\hypertarget{local_mod}{%
\subsection{Local subdomain sound field model}\label{local_mod}}

\newcommand{\captionone}{(Color online) Soundfield reconstruction via redundant subdomain decomposition in one spatial dimension (x-axis).
    The true signal $\mb p$ of length $N$ is partitioned in $S=N-N_s+1$ overlapping subdomains.
    All subdomains are of length $N_S$ and collected in $\mb P$.
    When the observed noisy signal $\mb{p}_\obs$ is partitioned (and $N_\obs<N$), $\mb P_\obs$ contains unknown values.
    Reconstruction is applied within each subdomain to estimate $\hat{\mb{P}}$.
    For each position, the corresponding elements in $\hat{\mb{P}}$ are averaged to yield the reconstructed sound field $\hat{\mb{p}}$.
}
\ifdefined\showfigures
\begin{figure}[!t]
    \includegraphics[width=\reprintcolumnwidth]{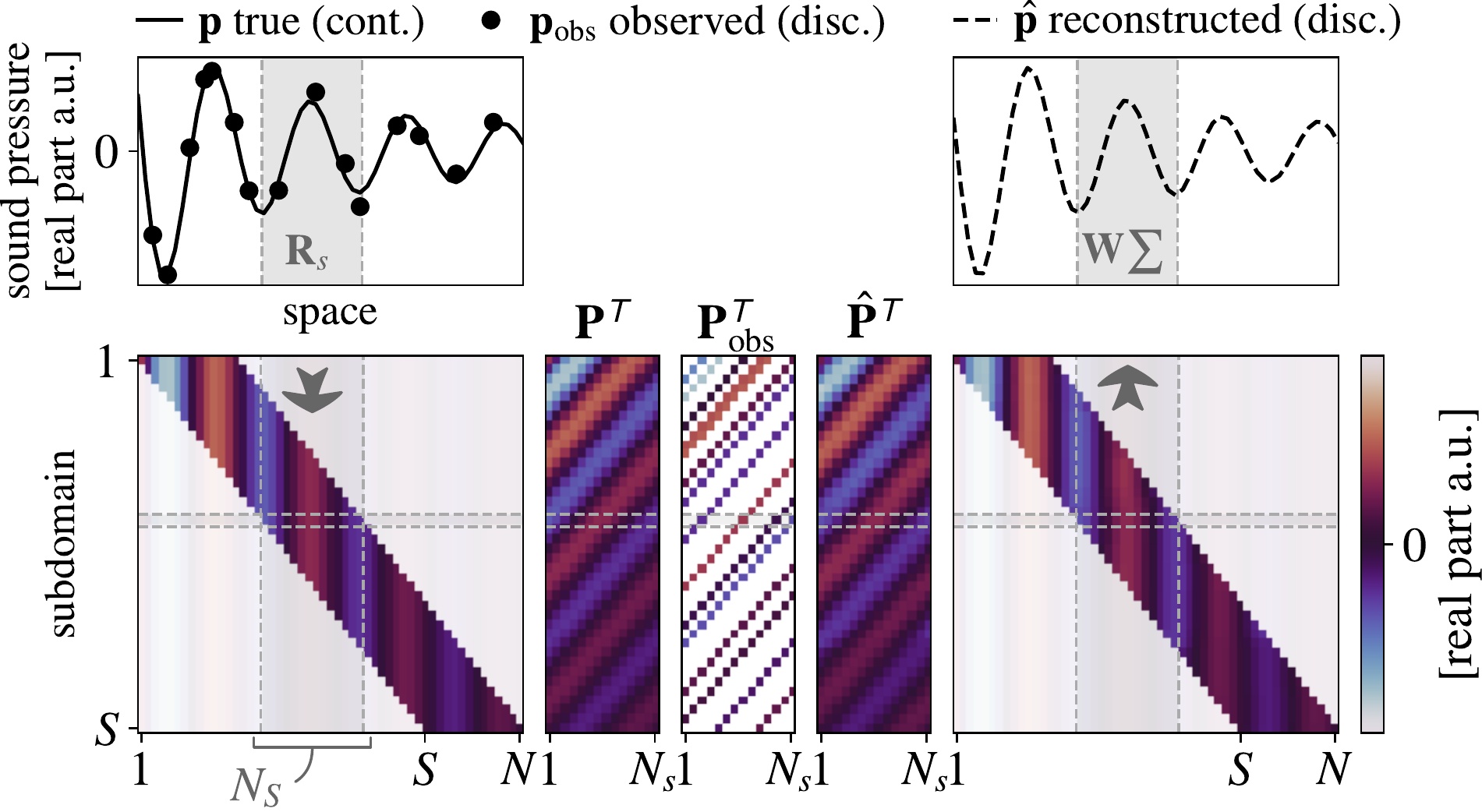}
    \caption{\captionone}
\end{figure}
\fi
\noindent

When reconstructing the sound field over large spatial domains (i.e.~much larger than the acoustic wavelength), it is useful to partition the global domain $\Omega$ into smaller, overlapping subdomains $\Omega_{sub}$. 
Correspondingly, the sound field $\mb p$ can be described by a collection of $S$ subdomain sound fields $\mb p_s \in \mathbb{C}^{N_s}$
\begin{equation}
    \mb P = [\mb R_1 \mb p \cdots \mb R_S \mb p] = [\mb p_1\cdots \mb p_S] \, , \ \mb P \in \mathbb{C}^{N_s \times S}.
    \label{eq:local}
\end{equation}
$\mb R_s \in \{0,1\}^{N_s \times N}$ is a binary extraction operator to select the $N_s$ positions contained in the $s\th$ subdomain $\Omega_s$.
The partitioning is illustrated on the left side of Fig.~1.
For simplicity, all subdomains within a sound field are considered to have the same extent, such that $N_s$ is constant.
Specifically, this study considers subdomains of extent one wavelength $\lambda$ in each dimension of the aperture.
Note that \eqref{eq:local} yields a redundant representation of the sound field if the subdomains overlap ($N_s S \geq N$).

A sound field can then be reconstructed within each subdomain of the partitioned observations $\mathbf P_\obs$, for example by applying the procedure in Sec.~\ref{global_rec} and finding coefficients $\hat{\mb x}_x$ via \eqref{eq:compressive_sensing} to estimate each $\hat{\mb p}_s$.
This yields the collection of reconstructed subdomain sound fields [see center right in Fig.~1].
\begin{equation}
    \hat{\mb P} = \mb H_{s} \hat{\mb X} \ ,
    \label{eq:local_rec}
\end{equation}
where $\mb H_{s}$ are the model functions at the desired positions within each local subdomain, for example $\mb H_s = \mb R_1 \mb H$, and 
\begin{equation}
    \hat{\mb X}= [\hat{\mb x}_1 \cdots \hat{\mb x}_S] \, , \ \hat{\mb X} \in \mathbb{C}^{M \times S}
    \label{eq:X}
\end{equation}
the estimated local coefficients.
The reconstructed field $\hat{\mb p}$ is reassembled from $\hat{\mb P}$ as the mean of overlapping subdomain representations [see right side of Fig.~1], 
\begin{equation}
    \hat{\mb{p}} = \mb{W} \, \sum_s \mb{R}_s^T \hat{\mb{p}_s} \, ,
    \label{eq:localtoglobal}
\end{equation}
where the diagonal matrix $\mb{W} = \text{diag}(\sum_s \mb{R}_s^T \mb 1_{N_s} )^{-1}$ normalizes by the spatial overlap of the subdomains and $\mb{1}_{N_s}$ denotes a vector of $N_s$ ones.

Such partitioning approaches counteract model mismatch, which occurs if a global model is suboptimal.
For example in the global sound field model of \eqref{eq:linear_model}, the chosen model functions in $\mb H$ might not span observations of a sound field across a large spatial domain.

Estimating independent coefficients for each subdomain allows for arbitrarily different wave components in each subdomain.
Such independent treatment of local representations disregards the similarity or even redundancy between sound fields in nearby or overlapping local subdomains.

The coefficients in the local subdomain sound field model can also be understood as a collection of coefficient maps $\mb x_m \in \mathbb{C}^S$, the rows in $\mb X = [\mb x\T_1 \cdots \mb x\T_M]\T$. 
The $m\th$ row in $\mb X$ contains a coefficient for the $m\th$ local model function across all $S$ subdomain locations.
When plane waves are used, the coefficient map $\mb x_m$ contains the spatial distribution of coefficients for the $m\th$ plane wave across subdomain representations.

\hypertarget{conv_mod}{%
\subsection{Convolutional sound field model}\label{conv_mod}}
\noindent


This study explores, how the spatial variations across $\mb x_m$, the rows of $\mb X$, can be taken into account for reconstruction, such that the global structure (or topology) of the sound field can be preserved.
For the remainder of the paper, we consider the reconstruction positions on a regular grid (containing also the measured positions as a subset).
Further, we assume subdomains of equal size and with full overlap, such that $S=N$, the number of subdomains equals the number of positions in $\mb p$ (circular boundary conditions).

The true sound field across an $L$-dimensional aperture can be rewritten as a sum of $M$ convolutions:
\begin{equation} 
    \mb{p} = \sum_{m=1}^M \, \mb{h}_m \ldconv \mb{x}_m
    \label{eq:csr} \ ,
\end{equation}
where $\mb{h}_m \in \mathbb{C}^{N_s}$ describes the $m\th$ local filter (the $m\th$ column of $\mb H_s$), for example a plane wave.
$\ldconv$ denotes a circular convolution along $L\in{1,2,3}$ spatial dimensions.
For example for $L=1$, the $n\th$ element of $\mb p$ is, 
\begin{align*}
    \mb{p}(n) =& \sum_m (\mb{h}_m \circledast \mb{x}_m)(n) \\
    =& \sum_m \left(\sum_{k=0}^{N_s} \mb h_m(k) \mb x_m \left( (N+n-k)\%N \right) \right) \ , 
\end{align*}
where $n=1,\ldots N$ and $\%$ is the modulo operation.

Compared to the collection of local sound fields in Sec.~\ref{local_mod}, this convolutional model reflects the spatial relations of coefficients and enables a joint analysis of all local representations in the global field.
For example, each subdomain representation can take the coefficients in nearby subdomains into account.
We propose to estimate the coefficients of \eqref{eq:csr} as
\begin{align}
    \hat{\mb X} &= \text{arg} \min_{\mb X} \; \frac 1 2
    \left\lVert \mb{M} \left( \sum_m \, \mb{h}_m \ldconv \mb{x}_m \right)  - \ \mb{p}_\obs\right\rVert_2^2 \label{eq:csc_mask_grad} \\
    &
    + \frac \mu 2 \,  \sum_m \sum_l \ \left\lVert \Delta_l \mb{x}_m \right\rVert_2^2 
    + \ \beta \sum_m \left\lVert \mb{x}_m \right\rVert_1
    \nonumber \ ,
\end{align}
The $\ell_1$ penalty promotes sparse coefficients, notably applied on the global coefficient vector.
A penalty on the spatial differences of the coefficients promotes smooth coefficient maps $\hat{\mb x}_m$, weighted by the regularization parameter $\mu$.
Specifically, $\Delta_l \mb x_m$ are the first order finite differences of the $m\th$ coefficient map along the $l\th$ dimension.
For example, when the considered aperture (and therefore $\mb x_m$) is one-dimensional, it is $\Delta \mb x_m = [x_{m1}-x_{m2}, x_{m2}-x_{m3} \cdots x_{mS}-x_{m1}]\T$.
Smooth coefficient maps $\hat{\mb x}_m$ enforce similarity between nearby and overlapping representations, which seems particularly suitable for sound fields.


\hypertarget{conv_admm}{%
\subsection{Convolutional reconstruction via ADMM}\label{conv_admm}}
\noindent


To reconstruct a sound field via the convolutional model, we solve \eqref{eq:csc_mask_grad} via the
\emph{alternating direction method of multipliers} (ADMM)\cite{boyd2010admm} and rewrite
\eqref{eq:csc_mask_grad} in matrix form
\begin{align}
    \hat{\mb x} = \text{arg} \min_{\mb x} \frac{1}{2}& \left\lVert \mb{MD_*} \mb{x} - \mb{p}_\obs \right\rVert_2^2  \label{eq:matrix_form} \\
    +& \frac \mu 2  \sum_m \sum_l \left\lVert \mb{G_*}_l \mb x_m \right\rVert_{2}^2
    + \beta \left\lVert \mb{x} \right\rVert_1
    \, , \nonumber
\end{align}
where $\mb x \in \mathbb{C}^{MN}$ are the stacked columns of $\mb X$ and
$\mb{D_*} \in \mathbb{C}^{N \times MN}$ is convolutional dictionary matrix such that $\mb{D_*} \mb x  = \sum_m \mb h_m \ldconv \mb x_m$ (e.g.~ for $L=1$, $\mb{D_*}$ is block-circulant with block $\mb H_s$).
$\mb{G_*}_l \mb x_m$ calculates the first order finite differences of the $m\th$ coefficient map along the $l\th$ dimension.
In the one-dimensional case, $\mb{G_*}$ is a circulant matrix with the first row $[1, -1, \mb 0_{1 \times S-2}]$, where $\mb 0_{1 \times S-2}$ is a row vector of $S-2$ zeros.

To solve \eqref{eq:matrix_form}, we split the variables and reformulate the joint problem \cite{wohlberg2016a,wohlberg2017a,wohlberg2018a} as
\begin{align}
   \underset{\mb {x,y_0,y_1}}{\text{minimize}} \quad &\frac{1}{2} \left\lVert \mb{M} \mb{y_0} - \mb{p}_\obs \right\rVert_2^2
    + \frac \mu 2 \sum_m \sum_l \left\lVert \mb{G_*}_l \mb{x} \right\rVert_{2}^2
    + \beta \left\lVert \mb{y_1} \right\rVert_1 
    \label{eq:admm_form}\\
    \text{subject to} \quad &\mb{Ax-y} = 0 \quad, \nonumber \\
    \text{where} \quad & \mb{A} = \begin{bmatrix} \mb{D_*} \\ \mb{I} \end{bmatrix}
    \quad \text{and} \quad 
    \mb{y} = \begin{bmatrix} \mb{y_0} \\ \mb{y_1} \end{bmatrix}
    . \nonumber
\end{align}
The ADMM steps in the $k\th$ iteration are
\begin{align}
    \mb{x}\kn &= \text{arg} \min_{\mb x} 
    \frac \mu 2 \sum_m \sum_l  \left\lVert \mb{G_*}_l \mb{x}_m \right\rVert_{2}^2
    \label{eq:xstep}\\
    &+ \frac \rho 2 \left\lVert \mb{A}\mb{x} - \mb{y}^k+ \mb{u}^k\right\rVert_2^2 
    \nonumber \\
    \mb{y}\kn &= \text{arg} 
    \min_{\mb y} \frac 1 2 \left\lVert \mb{M} \mb{y_0} - \ \mb{p}_\obs \right\rVert_2^2 \, + \beta \left\lVert \mb{y_1} \right\rVert_1
    \label{eq:ystep}\\
    &+ \frac \rho 2 \left\lVert \mb{Ax}\kn - \mb{y} + \mb{u}^k
        \right\rVert_2^2 \nonumber \\  
        \mb{u}\kn &= \mb{u}^k+ \mb{A x}\kn - \mb{y}\kn \ ,
    \label{eq:ustep}
\end{align}
where $\mb u = [ \mb{u_0}, \mb{u_1}]\T$ is the dual variable. The upper index $k$ indicates the state before the $k\th$ iteration (omitted further on for readability).
In spatial frequency domain, the convolutional matrices reduce to their transformed filters and \eqref{eq:xstep}
reduces to \cite{heide2015a}
\begin{equation}
    \left( \mu \mb{\tilde G}\H \mb{\tilde G} 
        + \rho \mb{\tilde D}\H \mb{\tilde D} \right)
    \mb{\tilde x}\kn = 
    \mb{\tilde D}\H \left(\mb{\tilde y_0} -
        \mb{\tilde u_0} \right) 
    + \left(\mb{\tilde y_1} -
        \mb{\tilde u_1} \right) \label{eq:xsolve} \, ,
\end{equation} 
which can be efficiently solved via the Sherman-Morrison formula \cite{wohlberg-2014-efficient,wohlberg2016b} and where $\tilde{\cdot}$ indicates frequency domain quantities.
Also, $\mb{\tilde{G}}\H\mb{\tilde{G}} = \sum_l \mb{\tilde{G}}_l\H\mb{\tilde{G}}_l$, where
$\mb{\tilde{G}}_l$ is the frequency transformed finite difference matrix in the $l\th$ dimension.\\
To align dimensions of $\mb{\tilde{D}}$ to the sound field $\mb{\tilde{p}}$, zero-padding is typically applied to the local filters $\mb h_m$ (corresponding to $[\mb H_s\T, \mb 0_{M \times N-N_s}]\T$, i.e. the first $M$ columns of $\mb D_*$).
Instead, we obtain $\mb{\tilde{D}}$ from plane waves functions evaluated over the complete sound field (i.e.~the global plane wave expansion $\mb H$).
\\
Equations \originaleqref{eq:ystep} and \originaleqref{eq:ustep} are separable, such that
\begin{align}
    \mb{y_0}\kn &= \text{arg} \min_{\mb{y_0}} \frac 1 2 \left\lVert \mb{M y_0} - \ \mb{p}_\obs \right\rVert_2^2     
        \label{eq:y0step}\\ 
        &+ \frac \rho 2 \left\lVert \mb{y_0} -
            \left(\mb{D_*x}\kn + \mb{u_0}\kz\right) \right\rVert_2^2 
    \nonumber \\
    \mb{y_1}\kn &= \text{arg} \min_{\mb{y_1}} \beta \left\lVert \mb{y_1} \right\rVert_1
        \label{eq:y1step}\\
        &+ \frac \rho 2 \left\lVert \mb{y_1} -
            \left(\mb{x}\kn  + \mb{u_1}\kz\right) \right\rVert_2^2 
        \nonumber\\
    \mb{u_0}\kn &= \mb{u_0}\kz+ \mb{D_* x}\kn - \mb{y_0}\kn
        \label{eq:u0step}\\
    \mb{u_1}\kn &= \mb{u_1}\kz+ \mb{x}\kn - \mb{y_1}\kn
        \label{eq:u1step} \ , 
\end{align}
where the $\mb{y_0}$ update \eqref{eq:y0step} has an efficient closed-form solution in frequency domain 
\begin{align}
    \mb{\tilde{y}_0}\kn &= 
    \left(\mb{\tilde{M}}\H \mb{\tilde{M}} + \rho \mb{I}\right)^{-1}
    \left(\mb{\tilde{M}\tilde{p}}_\obs + \rho \left(
            \mb{\tilde{D}\tilde{x}}\kn + \mb{\tilde{u}_0}\kz
        \right) \right) \ .
    \label{eq:y0solve}
\end{align}
$\mb{y_1}$ is updated by soft-thresholding, separable along the elements of $\mb{y_1}$,
\begin{align}
    \mb{y_1}\kn &= \mathcal{S}_{\beta/\rho}
    \left(\mb{x}\kn  + \mb{u_1}\kz\right)
    \label{eq:y1solve} \ ,
\end{align}
where $\mathcal{S}$ is the shrinkage operator 
\begin{align}
    \mathcal{S}_\alpha(\mb{z}) = \text{sign}(\mb{z}) \odot \text{max}(0, |\mb{z}|-\alpha) \ .
\end{align}

\hypertarget{eval}{%
    \subsection{Assessment of reconstructed sound fields}\label{eval}}
\noindent

To assess a reconstructed pressure field $\hat{\mb{p}}$, it is compared to the true field $\mb p$ in terms of the normalized mean square error \emph{NMSE} and the spatial similarity $C$.
The NMSE is
\begin{align}
    \text{NMSE} = 20 \log_{10} \left( \frac{\left\Vert\mb{\hat{p}} - \mb{p} \right\Vert_2 }{ \left\Vert \mb{p} \right\Vert_2 } \right) \ .
    \label{eq:nmse}
\end{align}
The spatial similarity $C$ is assessed as
\begin{align}
    C = \frac{\left| \mb{\hat{p}}^\text H \mb{p}\right|^2 }
    {
        \left(\mb{\hat{p}}^\text H \mb{\hat{p}} \right)
        \,
        \left(\mb{p}^\text H \mb{p}\right)
    } \ ,
    \label{eq:spatial_similarity}
\end{align}
such that $C=0$ indicates no similiarity and $C=1$ means the fields are indistinguishable.

\hypertarget{results}{%
    \section{Results}\label{results}}

To demonstrate the proposed approach, we reconstruct simulated and measured sound fields.
The proposed method is implemented with help of the \texttt{SPORCO} Python package \cite{wohlberg-2017-sporco}
and included as supplementary material.\footnote{}

\hypertarget{1dmonopole}{%
\subsection{Reconstruction along the radial distance from a monopole}\label{1dmonopole}}
\noindent

\newcommand{\captiontwo}{(Color online) Reconstruction of the sound field from a monopole (simulation), along a linear array in the radial dimension, placed at a distance of 0.5 $\lambda$.
    From top to bottom: normalized pressure and decimated measurements; reconstruction via convolutional model, independent local reconstructions and global plane waves; reconstruction errors. 
    Shown is the real part of the pressure in (a) and (b) and the absolute error in (c), normalized to the maximum pressure within the domain, arbitrary units.}
\ifdefined\showfigures
\begin{figure}[!t]
    \includegraphics[width=\reprintcolumnwidth]{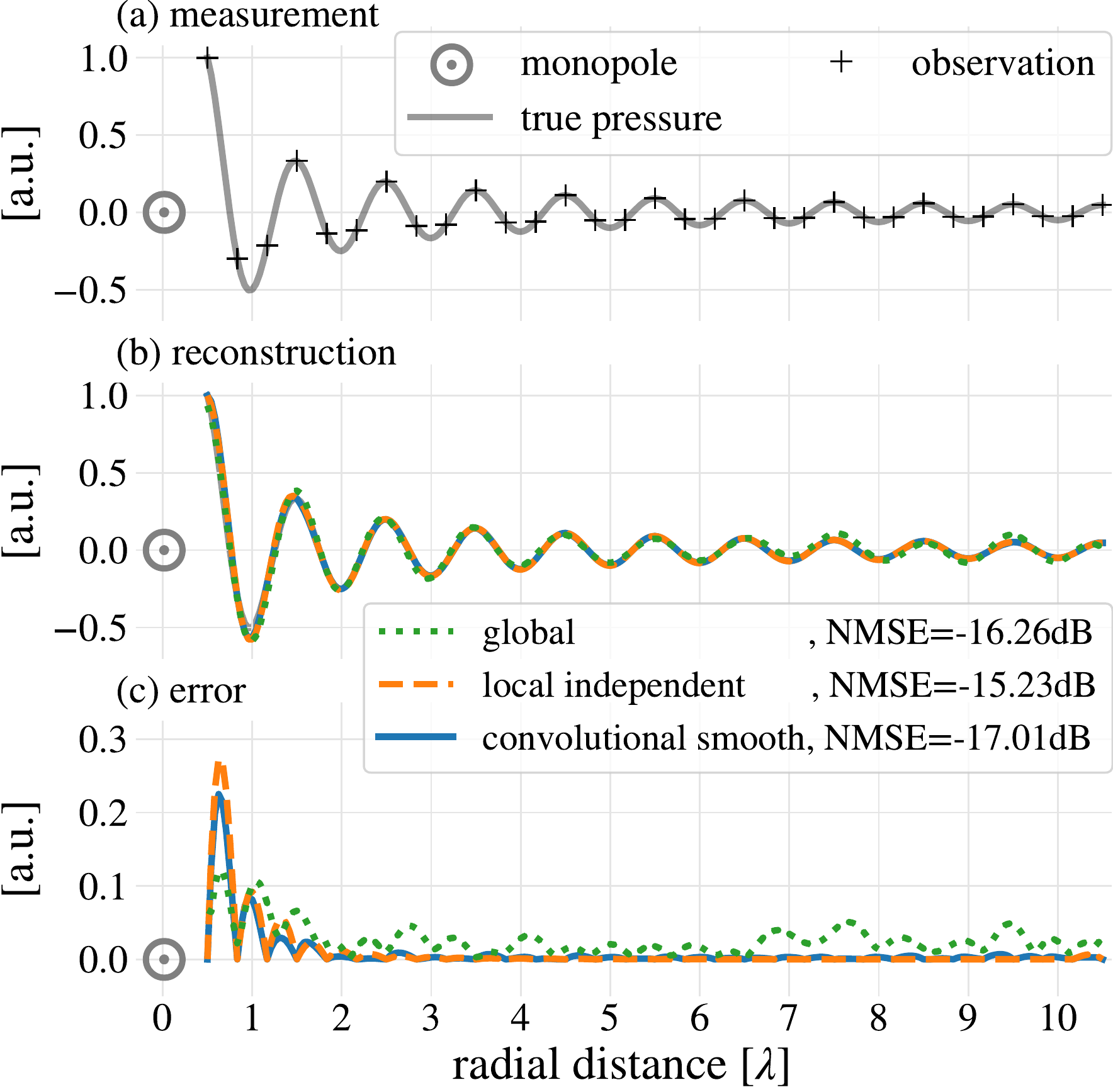}
    \caption{\captiontwo}
\end{figure}
\fi

The first experiment reconstructs the sound field by a monopole at the end of a linear microphone array, see Fig.~2(a).
The sound pressure is simulated radially across ten wavelengths ($\lambda$), spanning the near-field and the far-field of the monopole.
The linear array consists of 31 microphones at $0.5\,\lambda$ to $10.5\,\lambda$ radial distance and with spacing of $\lambda/3$.
Three reconstruction methods are compared:

\begin{enumerate}
    \setlength{\itemsep}{1pt}
    \setlength{\parskip}{0pt}
    \setlength{\parsep}{0pt}
    \item global plane waves with least squares (no regularization in \eqref{eq:compressive_sensing});
    \item local independent plane waves using compressive sensing, i.e. finding sparse representations via \eqref{eq:compressive_sensing} for each local partition separately, then overlap and average; 
    \item convolutional sparse plane waves with smooth coefficients via \eqref{eq:csc_mask_grad}.
\end{enumerate}

All three reconstruction methods use the same set of plane waves, with wavenumber $\|\mb k\|_2=\tfrac{2 \pi}{\lambda}$ and 21 incidence angles equally spaced along a semicircle $[0\cdots\pi]$. 
This experiment interpolates from observations with resultion $\lambda/3$ to a grid spacing of $\lambda/24$. 
The aperture of length $10\lambda$ contains $N=241$ reconstruction positions.
The local approaches operate use subdomains of size one wavelength.
In this example, each subdomain contains $N_s = 25$ reconstruction points and at most 4 observations.
Both sides of the domain are padded with $N_s-1$ zeros, to reduce artifacts from circular wrapping. 
The zero-padded domain has size $N'=N+2(N_s-1)=289$ and is partitioned in $S=N'$ fully overlapping subdomains.
After reconstruction, the sound field is cropped again to the original size $N$.
For comparison, the local approaches determine $MS = 21\times289$ coefficients compared to $M=21$ in the global model.

The reconstructions $\hat{\mb{p}}$ are shown in Fig.~2(b) and the error $\hat{\mb{p}} - \mb{p}$ in Fig.~2(c).
All methods yield good reconstructions.
The error is lowest for the smooth convolutional model, where smooth coefficients (plane wave magnitude and direction) among neighbouring representations are enforced.
Methods using locally variant plane wave coefficients are flexible enough to approximate all measurements (the error is zero at measurement positions). 
Still, they rely on sufficient measurements within a local partition to yield good predictions.
The global plane wave model can not conform to all measurements across the array, because of the mismatch between the radial decay of the sound field and propagating plane waves.

\newcommand{\captionthree}{(Color online) NMSE for reconstructions across a linear microphone array for varying distance between the monopole and array (see Fig.~2).}
\ifdefined\showfigures
\begin{figure}[!t]
    \includegraphics[width=0.85\reprintcolumnwidth]{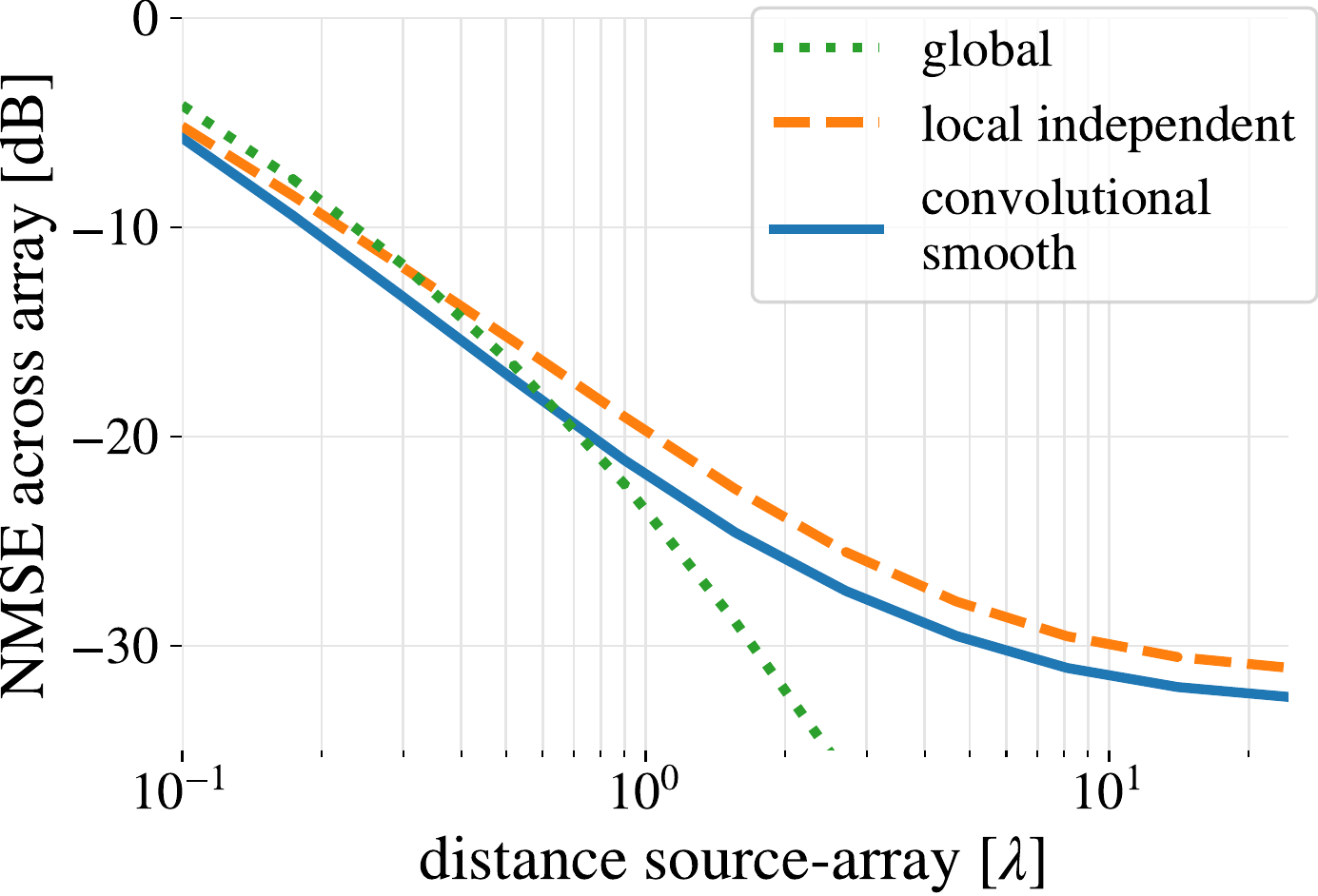}
    \caption{\captionthree}
\end{figure}
\fi

The experiment is repeated for varying distances between the monopole source and the microphone array. The normalized spatial mean squared error (\ref{eq:nmse}) is shown in Fig.~3.
When the array is close to the monopole, local representations improve reconstructions due to the field's high curvature and strong decay with distance.
The proposed smooth-convolutional approach gives the most accurate reconstructions up to 0.7$\lambda$.
The error decreases with distance for all methods, due to the less pronounced magnitude decay in the field.
The further the array is placed in far field, the more the true field approaches planar characteristics, such that also the global model fits to the observations well and yields the best predictions.
It is to note that in an application scenario, the reconstruction quality depends on many factors, such as the aperture size, curvature of wavefronts within the aperture, number and distribution of measurements available and not at least the evaluation criteria.

\hypertarget{2dmonoplane}{%
\subsection{2D reconstruction: monopole and plane wave}\label{2dmonoplane}}
\noindent
\label{sec:2dmonoplane}

\newcommand{\captionfour}{(Color online) 
Reconstructions of the sound pressure generated by a monopole interfering with a plane wave. Reconstructions from 100 (top row) and 300 microphones (bottom row).
Left to right:  measurements, reconstructions using global plane waves, independent local sparse representations, joint analysis with global sparsity, joint analysis with global sparsity and continuity, true reference. 
The aligned color scale is in dB SPL.}
\ifdefined\showfigures
\begin{figure*}[!t]
    \includegraphics[width=\textwidth]{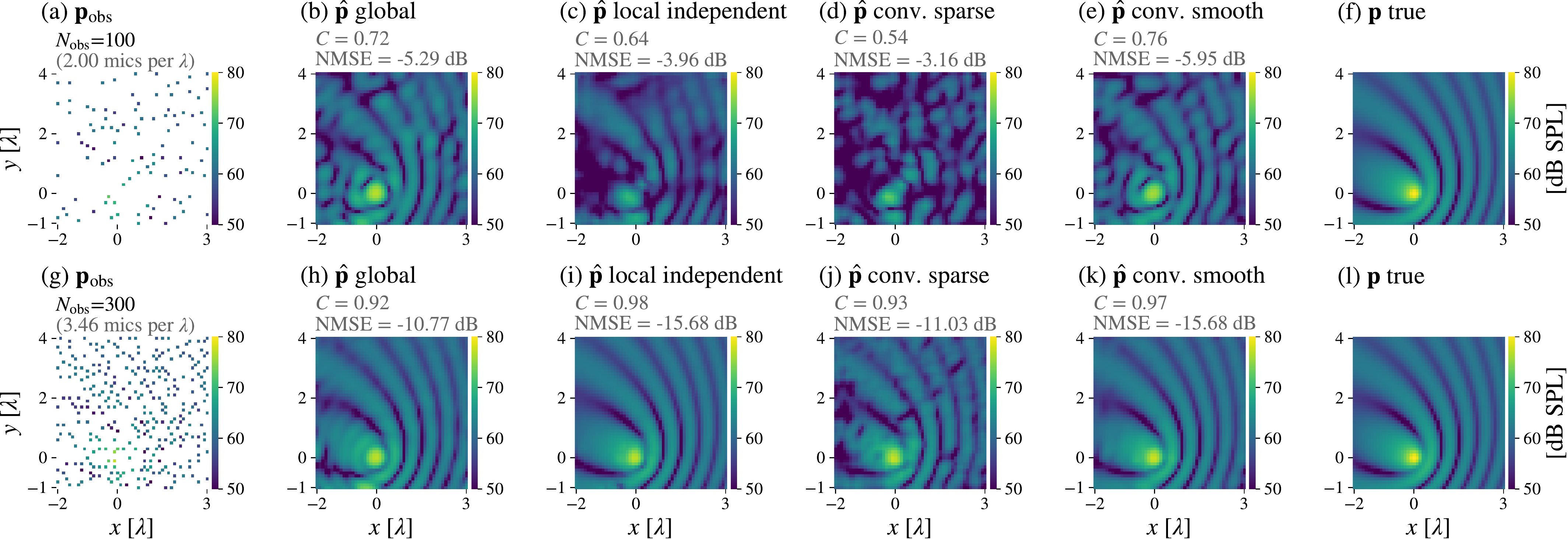}
    \caption{\captionfour}
\end{figure*}
\fi

The approach is tested for a two-dimensional aperture of size $(5\,\lambda)^2$ in the plane $z=0$.
The sound field is generated by interference of a monopole at (0,0,$\lambda$/8) with a plane wave propagating in $\mb k/k = (k_x,k_y,k_z)/k = (0.38, -0.76, 0.52)$.
Four plane wave reconstructions are tested:
\begin{enumerate}
    \setlength{\itemsep}{1pt}
    \setlength{\parskip}{0pt}
    \setlength{\parsep}{0pt}
    \item global plane waves with ridge regression ($\ell_2$-norm regularization in \eqref{eq:compressive_sensing}, parameter $\beta$ via leave-one-out cross-validation);
    \item local independent, sparse plane waves (solving \eqref{eq:compressive_sensing} via least-angle regression, regularization via leave-one-out cross-validation);
    \item convolutional sparse plane waves without smooth coefficients ($\mu=0$ in \eqref{eq:csc_mask_grad});
    \item convolutional sparse plane waves with smooth coefficients via \eqref{eq:csc_mask_grad}.
\end{enumerate}

All models use plane waves with propagation angles distributed in a fibonacci grid ($k_z\geq0$ hemisphere).
The global method i) uses $M$ = 1000 propagation angles, the local methods ii-iv) $M$ = 100 angles.
The reconstruction grid is regular with spacing of $d = \lambda/10$ between positions ($N = 51^2$).
The local subdomains and filters in methods ii)-iv) have size $\lambda^2$ (i.e. $N_s = (\lfloor \lambda / d \rfloor +1) ^ 2 = 11^2$ discrete positions).
For methods iii) and iv), the domain is zeropadded (with $\sqrt{N_s}-1$ in each direction) to avoid artifacts from circular convolutions ($N' = (\sqrt{N} + 2(\sqrt{N_s}-1))^2 = 3481$).
The regularization parameters are tuned to $\beta = 1 \times 10^{-5}$, $\mu = 1\times 10^{-3}$ (0 for iii)\,), $\rho = 1 \times 10^{-5}$, and the ADMM iterations are stopped after 500 iterations.
Note that this demonstration showcase exhibits a high signal to noise ratio. In less favourable conditions, the regularization would likely need to be adjusted.

The results are shown in Fig.~4.
For reconstructions from 100 microphones, the global and the proposed approach with smooth local coefficients capture the spatial phenomena better than the two other approaches ii) and iii), which only rely on local and global sparsity.
The spatially invariant (global) and slowly varying (proposed) models prescribe the necessary spatial structure to reconstruct from few measurements.
Both methods exploit measurements across a larger spatial range and capture the global structure of the sound field, which is critical in sparsely sampled scenarios (see Fig.~4(b) and (e) vs. (c-d).
When more microphones are available, spatially variant (local) models benefit from their flexibility to model complex sound fields.
The global approach exhibits artefacts around the monopole due to the strong decay and curvature of the wavefronts in this region.
The proposed approach balances local flexibility with global structure and yield good reconstructions in both cases.
Note that method iii) and the proposed method iv) apply the same shrinkage threshold $\beta/\rho$ to all local coefficients (see \eqref{eq:y1solve}).
Dynamic regularization could further improve the results, as it was observed for the independent local approach ii) (which uses cross-validation to find an optimal $\beta$ for each subdomain).

\newcommand{\captionfive}{(Color online) 
Particle velocity and intensity fields for the test using 300 microphones (second row of Fig.~4) to reconstruct the sound field by a monopole and an interfering plane wave (see Sec.~\ref{sec:2dmonoplane}).
From top to bottom: true field, reconstructions using a global, local independent and convolutional smooth model.
Vector fields for particle velocity $\Re\{\mb u_{xy}\}$ and intensity $\Re\{\mb p \odot \mb u_{xy}^*\}$. 
For readability, vector norms are clipped, $|\mb u|$ to $150\,\mu\text{ms}^{-1}$ and $|\mb I|$ to $2\,\mu\text{Wm}^{-2}$, and only every fourth intensity vector is shown.
Color indicates particle velocity and intensity levels in dB, where $u_{ref}=50\,$nms$^{-1}$ and $I_{ref}=1\,$pWm$^{-2}$, and bilinear interpolation is used for readability.
}
\ifdefined\showfigures
\begin{figure}
    \includegraphics[width=\reprintcolumnwidth]{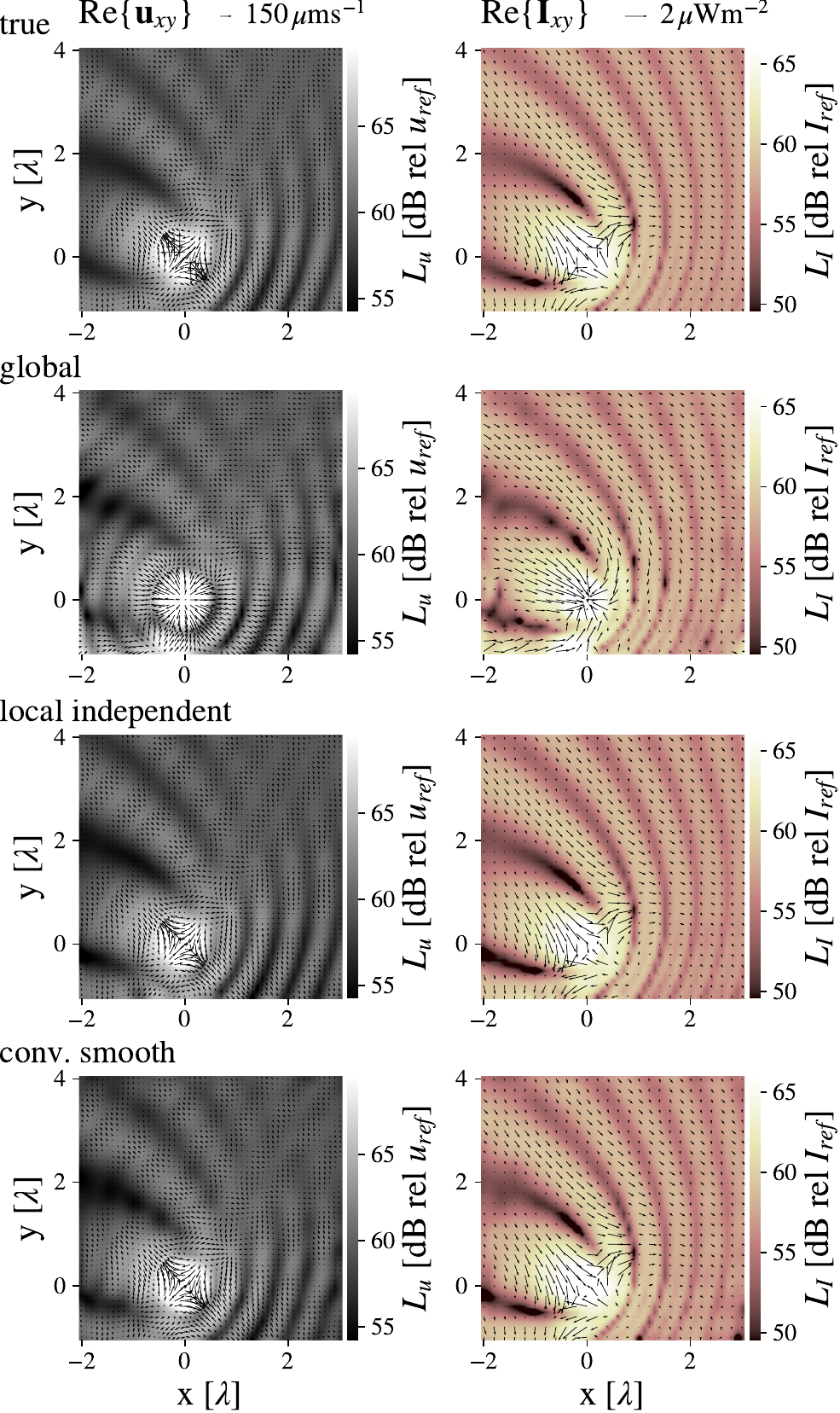}
    \caption{\captionfive}
\end{figure}
\fi

The particle velocity and sound intensity xy-vector fields of the reconstructions from 300 microphones (second row Fig.~4) are shown in Fig.~5.
Global representations can not conform to the drastic spatial variations close to the monopole, all particle velocity vectors point outwards from $(x,y)=(0,0)$.
Local approaches recover the fine structure of particle velocity and intensity also around the monopole. 

\hypertarget{classroom}{%
\subsection{Experimental reconstruction with real data: classroom measurement}\label{classroom}}
\noindent

\newcommand{\captionsix}{(Color online)
    Furnished classroom (room 019 in DTU building 352, Lyngby, Denmark) with absorbing ceiling and wooden floor (left, picture from Ref.~\onlinecite{hahmann2021spatial}) and robotic arm (right).
}
\ifdefined\showfigures
\begin{figure}[!t]
    \includegraphics[width=\reprintcolumnwidth]{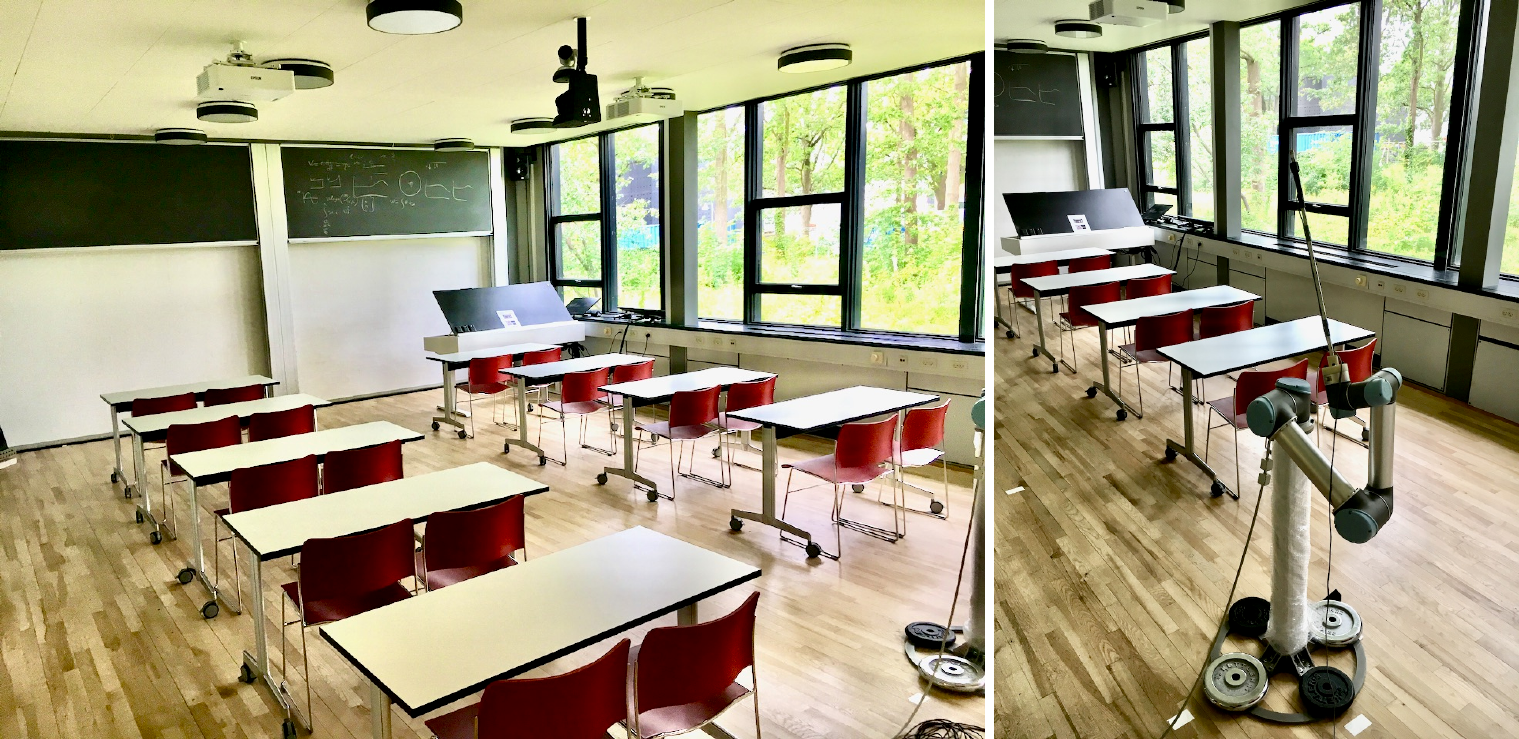}
    \caption{\captionsix}
\end{figure}
\fi

\newcommand{\captionseven}{(Color online)
Reconstructions of the sound pressure field in a classroom, from 98 (top row) and 295 microphones (bottom row).
Left to right: measurements, reconstructions using global plane waves, independent local sparse representations, joint analysis with global sparsity, joint analysis with global sparsity and continuity, true reference.
The aligned color scale is in dB, relative to the spatial mean of the squared true pressure field, $<\mathbf{p}_{true}^2>$.}
\ifdefined\showfigures
\begin{figure*}[!t]
    \includegraphics[width=\textwidth]{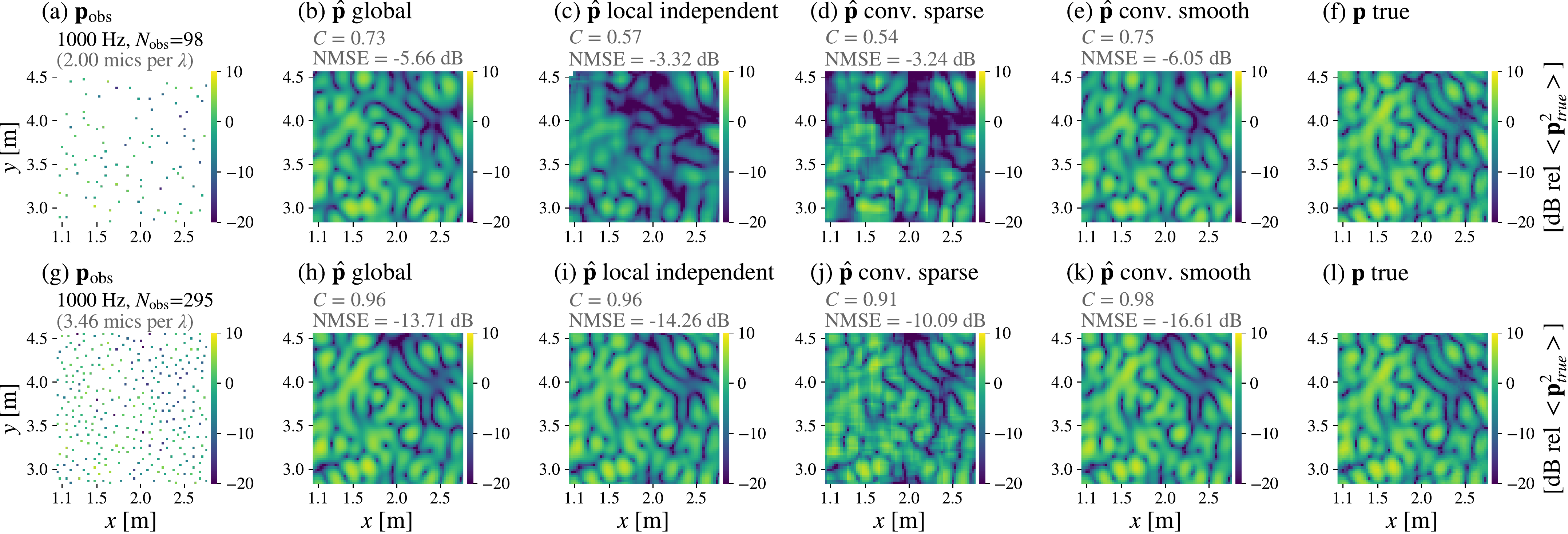}
    \caption{\captionseven}
\end{figure*}
\fi

The same methods i)-iv) from Sec.~\ref{sec:2dmonoplane} (and parameters) are used to reconstruct the enclosed reverberant sound field in a classroom (DTU building 352, Lyngby, Denmark), shown in Fig.~6.
The room dimensions are $(l_x, l_y, l_z) = (6.63, 9.45, 2.97)$\,m, the reverberation time approx.~$T_{60} = 0.5$\,s, the Schr\"oder frequency $f_{S} \approx 240$\,Hz. 
The room is furnished and its walls are somewhat irregular, with wooden floor, scattering elements on the walls and absorbing ceiling.
A loudspeaker (BM6, Dynaudio, Skanderborg, Denmark) placed in a room corner was used to excite the room with 10\,s logarithmic sweeps from 20\,Hz to 20\,kHz.
A total of 4761 frequency responses were measured using a robotic arm (UR5, Universal Robots, Odense, Denmark) with a 1/2 inch free field condenser microphone (Br\"uel\&Kj\"aer, N\ae rum, Denmark).
The positions are distributed over a $1.7 \times 1.7$\,m$^2$ planar aperture with $N = 69^2$ positions on a regular grid with 2.5\,cm spacing.
We refer the reader to \onlinecite{hahmann2021spatial} for more information on the room and measurements and to \onlinecite{hahmann2021b} for the dataset.

The sound field in the classroom is reconstructed at 1000\,Hz from 98 and 295 measurements, distributed with uniform probability (and a minimum distance of 7\,cm) across the aperture.
Reconstructions using i-iv) and measured reference (``true'') of the classroom sound field are shown in Fig.~7 for the two cases with $N_\obs$ = 98 and 295. 
To reconstruct from few measurements, it is necessary to capture the global structure of the sound field, as in the global and the proposed approach (Fig.~7(b) and (d)).
Models based on local representations conform easier to many measurements due to their higher number of coefficients (Fig.~7(i-k)).
Specifically in this test, subdomains of size $\lambda^2$ contain $N_s = 24^2$ discrete reconstruction positions, such that local models use a total of ii) $MN = 476100$ and with padding iii+iv) $MN' = M (\sqrt{N} + 2(\sqrt{N_s}-1))^2 = 1322500$ coefficients compared to 1000 in the global model.
The proposed approach combines both local flexibility and joint global analysis.
As a consequence, it yields the highest similarity and lowest reconstruction errors when compared to the measured field.

The benefit of smooth coefficients shows when comparing the two convolutional approaches.
Both seek a sparse approximation of the measurements, but spatial continuity is required to reconstruct sound fields successfully via local representations.
Also the local independent approach yields smooth reconstructions, namely by averaging of overlapping partitions.
However, a joint analysis of nearby representations is needed to align nearby coefficients and hence, indirectly exploit nearby measurements. 
In this study, the sparsity constraint enables feasible reconstructions also when only few measurements are available within a local subdomain and the inverse problem is severely underdetermined.
As such, it is not the goal to represent the sound field using the fewest number of coefficients, but local sparsity is a means of exploiting the available measurements.

\newcommand{\captioneight}{(Color online)
Reconstruction error (NMSE) of the sound field in a classroom across frequency from (a) 80, (b) 160 and (c) 320 microphones.
Shown are NMSE mean $\pm$ one standard deviation, obtained from 12 reconstructions from pseudo-randomly distributed measurements.
The vertical line indicates the frequency, where the average distance between measurements is $\overline{d_m} = \lambda/2$.}
\ifdefined\showfigures
\begin{figure}[!t]
    \includegraphics[width=\reprintcolumnwidth]{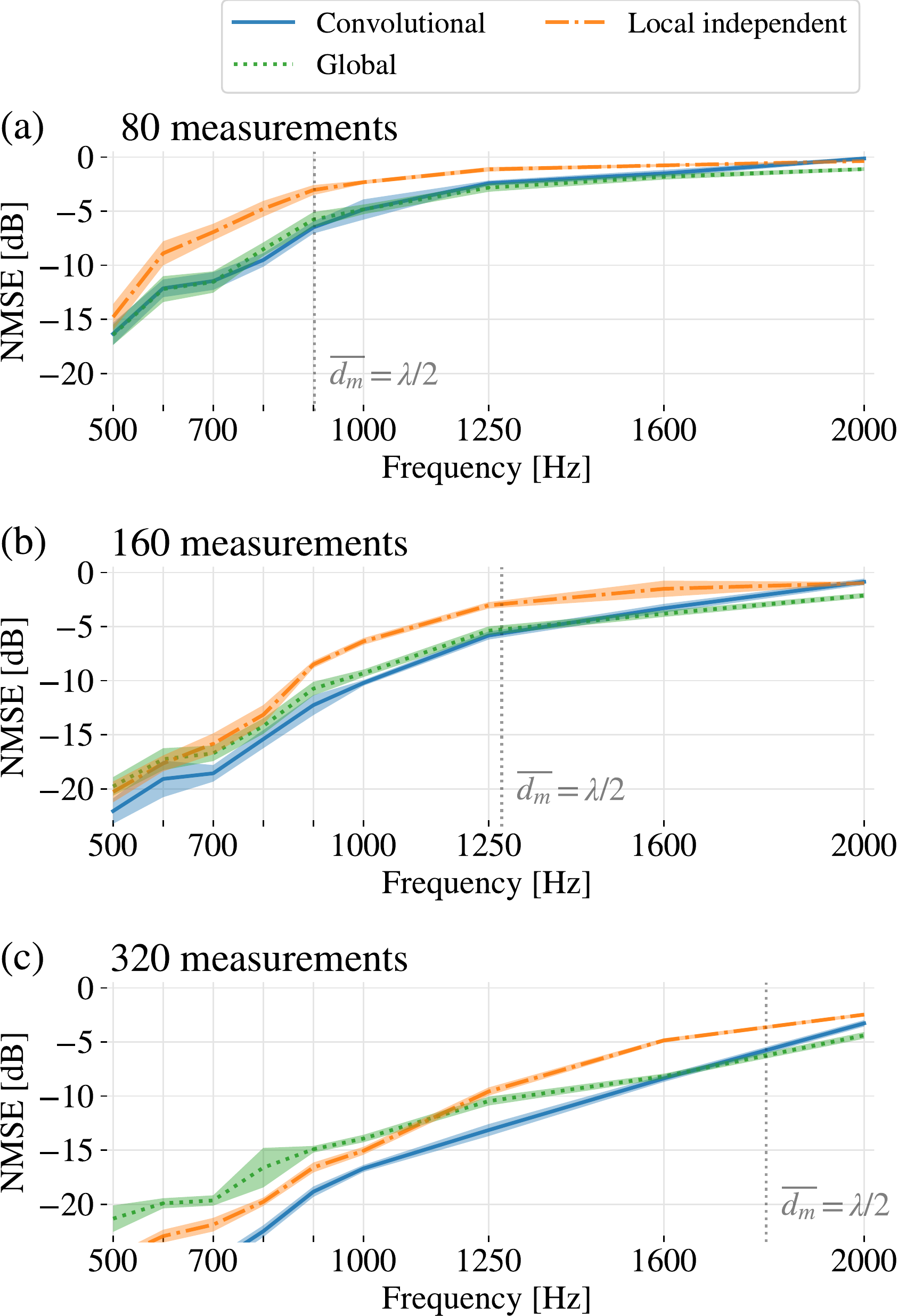}
    \caption{\captioneight}
\end{figure}
\fi

The experiment in Fig.~7 is extended to other frequencies and reconstruct the sound field in the classroom from 500\,Hz to 2\,kHz using a fixed number of microphones.
The NMSE results in Fig.~8 show that the proposed approach yields good reconstructions when average distance between measurements is lower than $\lambda/2$.
The proposed approach yields significant improvement with errors close to (see Fig.~8(a)), or lower than global plane waves (for sufficient measurements, see Fig.~8(b,c)).

    






\hypertarget{conclusion}{%
    \section{Conclusion}\label{conclusion}}
\noindent

This study formulates a sound field model as a spatial convolution between a global coefficient map and local plane wave filters.
This model leads to a joint analysis of all local representations, while keeping their spatial relation (and thereby the global structure of the field) intact.
By penalizing the spatial differences of plane wave coefficients, continuity between neighboring representations is enforced in terms of amplitude and direction of the plane waves.
In this way, each local representation has to be consistent with its neighbours, and can therefore utilise nearby observations.
The experiments indicate that the proposed approach both conforms to complex spatial sound fields and also preserves the global structure of the sound field.
Compared to other local models using locally sparse coding in terms of plane waves, the proposed approach attains better reconstructions of sound fields when few measurements are available.
When measurements are very scarcely distributed, an expansion of the entire global field in terms of plane waves yields the best reconstructions.
However, when sufficient measurements are available, the experiments indicate that local representation models conform best to fields of higher complexity.
This is shown for the reconstruction of the sound pressure, as well as for the reconstruction of particle velocity and sound intensity vector fields, where the improvements are even more substantial.

\acknowledgments{This work is funded by VILLUM Fonden through VILLUM Young Investigator grant number 19179 for the project ‘Large-scale Acoustic Holography’.}


\begin{thebibliography}{10}
\def\enquote#1,{``#1,''}
\def\enxquote#1{``#1''}
\expandafter\ifx\csname url\endcsname\relax
  \def\url#1{\texttt{#1}}\fi
\expandafter\ifx\csname urlprefix\endcsname\relax\def\urlprefix{URL }\fi
\providecommand{\bibinfo}[2]{#2}
\def\plainquote#1{``#1''}
\providecommand{\noopsort}[1]{}
\providecommand{\switchargs}[2]{#2#1}
\providecommand{\dourl}[1]{\href{http://#1}{\nolinkurl{#1}}}
  \def\eatspace #1{#1}

\bibitem{jacobsen2010a}
\bibinfo{author}{F.~Jacobsen} and \bibinfo{author}{E.~Tiana~Roig},
  \enquote{\bibinfo{title}{Measurement of the sound power incident on the walls
  of a reverberation room with near field acoustic holography}},
  \bibinfo{journal}{Acustica United w. Acta Acustica} \textbf{96}(1),
  \bibinfo{pages}{76--81} (\bibinfo{year}{2010}).

\bibitem{verburg2018a}
\bibinfo{author}{S.~A. Verburg} and \bibinfo{author}{E.~Fernandez-Grande},
  \enquote{\bibinfo{title}{Reconstruction of the sound field in a room using
  compressive sensing}},  \bibinfo{journal}{J. Acoust. Soc. Am.}
  \textbf{143}(6), \bibinfo{pages}{3770--3779} (\bibinfo{year}{2018}).

\bibitem{haneda1999a}
\bibinfo{author}{Y.~Haneda}, \bibinfo{author}{Y.~Kaneda}, and
  \bibinfo{author}{N.~Kitawaki},
  \enquote{\bibinfo{title}{Common-acoustical-pole and residue model and its
  application to spatial interpolation and extrapolation of a room transfer
  function}},  \bibinfo{journal}{IEEE Trans. Sp. Audio Proc.} \textbf{7}(6),
  \bibinfo{pages}{709--717} (\bibinfo{year}{1999}).

\bibitem{mignot2013a}
\bibinfo{author}{R.~Mignot}, \bibinfo{author}{L.~Daudet}, and
  \bibinfo{author}{F.~Ollivier}, \enquote{\bibinfo{title}{Room reverberation
  reconstruction: Interpolation of the early part using compressed sensing}},
  \bibinfo{journal}{IEEE/ACM Trans. Audio, Speech, Lang. Process.}
  \textbf{21}(11), \bibinfo{pages}{6562745, 2301--2312} (\bibinfo{year}{2013}).

\bibitem{mignot2014a}
\bibinfo{author}{R.~Mignot}, \bibinfo{author}{G.~Chardon}, and
  \bibinfo{author}{L.~Daudet}, \enquote{\bibinfo{title}{Low frequency
  interpolation of room impulse responses using compressed sensing}},
  \bibinfo{journal}{IEEE/ACM Trans. Audio, Speech, Lang. Process.}
  \textbf{22}(1), \bibinfo{pages}{205--216} (\bibinfo{year}{2014}).

\bibitem{nolan2019a}
\bibinfo{author}{M.~Nolan}, \bibinfo{author}{S.~A. Verburg},
  \bibinfo{author}{J.~Brunskog}, and \bibinfo{author}{E.~Fernandez-Grande},
  \enquote{\bibinfo{title}{Experimental characterization of the sound field in
  a reverberation room}},  \bibinfo{journal}{J. Acoust. Soc. Am.}
  \textbf{145}(4), \bibinfo{pages}{2237--2246} (\bibinfo{year}{2019}).

\bibitem{witew2017a}
\bibinfo{author}{I.~B. Witew}, \bibinfo{author}{M.~Vorländer}, and
  \bibinfo{author}{N.~Xiang}, \enquote{\bibinfo{title}{Sampling the sound field
  in auditoria using large natural-scale array measurements}},
  \bibinfo{journal}{J. Acoust. Soc. Am.} \textbf{141}(3),
  \bibinfo{pages}{EL300--EL306} (\bibinfo{year}{2017}).

\bibitem{brandao2022a}
\bibinfo{author}{E.~Brand\~ao} and \bibinfo{author}{E.~Fernandez-Grande},
  \enquote{\bibinfo{title}{Analysis of the sound field above finite absorbers
  in the wave-number domain}},  \bibinfo{journal}{J. Acoust. Soc. Am.}
  \textbf{151}(5), \bibinfo{pages}{3019--3030} (\bibinfo{year}{2022}).

\bibitem{heuchel2020a}
\bibinfo{author}{F.~M. Heuchel}, \bibinfo{author}{D.~Caviedes-Nozal},
  \bibinfo{author}{J.~Brunskog}, and \bibinfo{author}{F.~T. Agerkvist},
  \enquote{\bibinfo{title}{Large-scale outdoor sound field control}},
  \bibinfo{journal}{J. Acoust. Soc. Am.} \textbf{148}(4),
  \bibinfo{pages}{2392--2402} (\bibinfo{year}{2020}).

\bibitem{caviedes2019a}
\bibinfo{author}{D.~Caviedes-Nozal}, \bibinfo{author}{F.~M. Heuchel},
  \bibinfo{author}{J.~Brunskog}, \bibinfo{author}{N.~A.~B. Riis}, and
  \bibinfo{author}{E.~Fernandez-Grande}, \enquote{\bibinfo{title}{A bayesian
  spherical harmonics source radiation model for sound field control}},
  \bibinfo{journal}{J. Acoust. Soc. Am.} \textbf{146}(5),
  \bibinfo{pages}{3425--3435} (\bibinfo{year}{2019}).

\bibitem{heuchel2018a}
\bibinfo{author}{F.~M. Heuchel}, \bibinfo{author}{E.~Fernandez-Grande},
  \bibinfo{author}{F.~T. Agerkvist}, and \bibinfo{author}{E.~Shabalina},
  \enquote{\bibinfo{title}{Active room compensation for sound reinforcement
  using sound field separation techniques}},  \bibinfo{journal}{J. Acoust. Soc.
  Am.} \textbf{143}(3), \bibinfo{pages}{1346--1354} (\bibinfo{year}{2018}).

\bibitem{betlehem2005a}
\bibinfo{author}{T.~Betlehem} and \bibinfo{author}{T.~D. Abhayapala},
  \enquote{\bibinfo{title}{Theory and design of sound field reproduction in
  reverberant rooms}},  \bibinfo{journal}{J. Acoust. Soc. Am.} \textbf{117}(4),
  \bibinfo{pages}{2100--2111} (\bibinfo{year}{2005}).

\bibitem{moller2019a}
\bibinfo{author}{M.~B. Møller}, \bibinfo{author}{J.~K. Nielsen},
  \bibinfo{author}{E.~Fernandez-Grande}, and \bibinfo{author}{S.~K. Olesen},
  \enquote{\bibinfo{title}{On the influence of transfer function noise on sound
  zone control in a room}},  \bibinfo{journal}{IEEE/ACM Trans. Audio, Speech,
  Lang. Process.} \textbf{27}(9), \bibinfo{pages}{1405--1418}
  (\bibinfo{year}{2019}).

\bibitem{borrel2021a}
\bibinfo{author}{N.~Borrel-Jensen}, \bibinfo{author}{A.~P. Engsig-Karup}, and
  \bibinfo{author}{C.-H. Jeong}, \enquote{\bibinfo{title}{Physics-informed
  neural networks for one-dimensional sound field predictions with
  parameterized sources and impedance boundaries}},  \bibinfo{journal}{JASA
  Express Letters} \textbf{1}(12), \bibinfo{pages}{122402}
  (\bibinfo{year}{2021}).

\bibitem{tylka2017a}
\bibinfo{author}{J.~G. Tylka} and \bibinfo{author}{E.~Y. Choueiri},
  \enquote{\bibinfo{title}{Evaluation of techniques for navigation of
  higher-order ambisonics}},  \bibinfo{journal}{J. Acoust. Soc. Am.}
  \textbf{141}(5), \bibinfo{pages}{3511--3511} (\bibinfo{year}{2017}).

\bibitem{tylka2020a}
\bibinfo{author}{J.~G. Tylka} and \bibinfo{author}{E.~Y. Choueiri},
  \enquote{\bibinfo{title}{Fundamentals of a parametric method for virtual
  navigation within an array of ambisonics microphones}},  \bibinfo{journal}{J.
  Audio Eng. Soc.} \textbf{68}(3), \bibinfo{pages}{120--137}
  (\bibinfo{year}{2020}).

\bibitem{winter2014a}
\bibinfo{author}{F.~Winter}, \bibinfo{author}{F.~Schultz}, and
  \bibinfo{author}{S.~Spors}, \enquote{\bibinfo{title}{Localization properties
  of data-based binaural synthesis including translatory head-movements}},
  \bibinfo{journal}{Proceedings of Forum Acusticum} \textbf{2014-}
  (\bibinfo{year}{2014}).

\bibitem{schultz2013data}
\bibinfo{author}{F.~Schultz} and \bibinfo{author}{S.~Spors},
  \enquote{\bibinfo{title}{Data-based binaural synthesis including rotational
  and translatory head-movements}}, in \emph{\bibinfo{booktitle}{Audio Eng.
  Soc. Conf.: Sound Field Control - Eng. and Percep.}} (\bibinfo{year}{2013}).

\bibitem{fernandez2021a}
\bibinfo{author}{E.~Fernandez-Grande}, \bibinfo{author}{D.~Caviedes-Nozal},
  \bibinfo{author}{M.~Hahmann}, \bibinfo{author}{X.~Karakonstantis}, and
  \bibinfo{author}{S.~A. Verburg}, \enquote{\bibinfo{title}{Reconstruction of
  room impulse responses over extended domains for navigable sound field
  reproduction}}, in \emph{\bibinfo{booktitle}{Proceed. Int. Conf. Immers. 3D
  Audio}}, \bibinfo{publisher}{IEEE} (\bibinfo{year}{2021}), p.
  \bibinfo{pages}{8 pp.}

\bibitem{jacobsen2013a}
\bibinfo{author}{F.~Jacobsen} and \bibinfo{author}{P.~M. Juhl},
  \emph{\bibinfo{title}{Fundamentals of general linear acoustics}}
  (\bibinfo{publisher}{Wiley}, \bibinfo{address}{London},
  \bibinfo{year}{2013}).

\bibitem{moiola2011a}
\bibinfo{author}{A.~Moiola}, \bibinfo{author}{R.~Hiptmair}, and
  \bibinfo{author}{I.~Perugia}, \enquote{\bibinfo{title}{Plane wave
  approximation of homogeneous helmholtz solutions}},
  \bibinfo{journal}{Zeitschrift Fur Angewandte Mathematik Und Physik}
  \textbf{62}(5), \bibinfo{pages}{809--837} (\bibinfo{year}{2011}).

\bibitem{pezzoli2022sparsity}
\bibinfo{author}{M.~Pezzoli}, \bibinfo{author}{M.~Cobos}, 
\bibinfo{author}{F.~Antonacci}, and \bibinfo{author}{A.~Sarti}, 
\enquote{\bibinfo{title}{Sparsity-based sound field separation in the spherical harmonics domain}}, in
\emph{\bibinfo{booktitle}{IEEE Int. Conf. Acoust. Sp. Sig. Process. (ICASSP)}}
(\bibinfo{year}{2022}), pp. \bibinfo{pages}{1051--1055}.


\bibitem{tylka2020b}
\bibinfo{author}{J.~G. Tylka} and \bibinfo{author}{E.~Y. Choueiri},
  \enquote{\bibinfo{title}{Performance of linear extrapolation methods for
  virtual sound field navigation}},  \bibinfo{journal}{J. Audio Eng. Soc.}
  \textbf{68}(3), \bibinfo{pages}{138--156} (\bibinfo{year}{2020}).

\bibitem{elad2006a}
\bibinfo{author}{M.~Elad} and \bibinfo{author}{M.~Aharon},
  \enquote{\bibinfo{title}{Image denoising via sparse and redundant
  representations over learned dictionaries}},  \bibinfo{journal}{IEEE Trans.
  Image Process.} \textbf{15}(12), \bibinfo{pages}{3736--3745}
  (\bibinfo{year}{2006}).

\bibitem{markovic2016a}
\bibinfo{author}{D.~Markovic}, \bibinfo{author}{L.~Bianchi},
  \bibinfo{author}{S.~Tubaro}, and \bibinfo{author}{A.~Sarti},
  \enquote{\bibinfo{title}{Extraction of acoustic sources through the
  processing of sound field maps in the ray space}},
  \bibinfo{journal}{IEEE/ACM Trans. Audio, Speech, Lang. Process.}
  \textbf{24}(12), \bibinfo{pages}{2481--2494} (\bibinfo{year}{2016}).

\bibitem{markovic2013a}
\bibinfo{author}{D.~Markovic}, \bibinfo{author}{F.~Antonacci},
  \bibinfo{author}{A.~Sarti}, and \bibinfo{author}{S.~Tubaro},
  \enquote{\bibinfo{title}{Soundfield imaging in the ray space}},
  \bibinfo{journal}{IEEE/ACM Trans. Audio, Speech, Lang. Process.}
  \textbf{21}(12), \bibinfo{pages}{2493--2505} (\bibinfo{year}{2013}).

\bibitem{hahmann2021spatial}
\bibinfo{author}{M.~Hahmann}, \bibinfo{author}{S.~A. Verburg}, and
  \bibinfo{author}{E.~Fernandez-Grande}, \enquote{\bibinfo{title}{Spatial
  reconstruction of sound fields using local and data-driven functions}},
  \bibinfo{journal}{J. Acoust. Soc. Am.} \textbf{150}(6),
  \bibinfo{pages}{4417--4428} (\bibinfo{year}{2021}).

\bibitem{jin2017a}
\bibinfo{author}{C.~Jin}, \bibinfo{author}{F.~Antonacci}, and
  \bibinfo{author}{A.~Sarti}, \enquote{\bibinfo{title}{Ray space analysis with
  sparse recovery}}, in \emph{\bibinfo{booktitle}{2017 IEEE Works. Appl. Si.
  Process. Aud. Acous. (WASPAA)}} (\bibinfo{year}{2017}), pp.
  \bibinfo{pages}{239--243}.

\bibitem{yu2021upscaling}
\bibinfo{author}{S.~Yu}, \bibinfo{author}{C.~Jin},
  \bibinfo{author}{F.~Antonacci}, and \bibinfo{author}{A.~Sarti},
  \enquote{\bibinfo{title}{Sparse recovery beamforming and upscaling in the ray
  space}}, in \emph{\bibinfo{booktitle}{IEEE Int. Conf. Acoust. Sp. Sig.
  Process. (ICASSP)}} (\bibinfo{year}{2021}), pp. \bibinfo{pages}{776--780}.

\bibitem{morse1944a}
\bibinfo{author}{P.~Morse} and \bibinfo{author}{R.~Bolt},
  \enquote{\bibinfo{title}{Sound waves in rooms}},  \bibinfo{journal}{Reviews
  of Modern Physics} \textbf{16}(2), \bibinfo{pages}{0069--0150}
  (\bibinfo{year}{1944}).

\bibitem{schroeder1954a}
\bibinfo{author}{M.~Schr\"oder},
  \enquote{\bibinfo{title}{Eigenfrequenzstatistik und anregungsstatistik in r\"
  aumen - modellversuche mit elektrischen wellen}},
  \bibinfo{journal}{Acustica} \textbf{4}(4), \bibinfo{pages}{456--468}
  (\bibinfo{year}{1954}).

\bibitem{pierce1981a}
\bibinfo{author}{A.~D. Pierce}, \emph{\bibinfo{title}{Acoustics. An
  introduction to its physical principles and applications}}
  (\bibinfo{publisher}{McGraw-Hill}, \bibinfo{address}{New York},
  \bibinfo{year}{1981}).

\bibitem{papyan2017a}
\bibinfo{author}{V.~Papyan}, \bibinfo{author}{J.~Sulam}, and
  \bibinfo{author}{M.~Elad}, \enquote{\bibinfo{title}{Working locally thinking
  globally: Theoretical guarantees for convolutional sparse coding}},
  \bibinfo{journal}{IEEE Trans. Signal Process.} \textbf{65}(21),
  \bibinfo{pages}{7997798, 5687--5701} (\bibinfo{year}{2017}).

\bibitem{wohlberg2018a}
\bibinfo{author}{B.~Wohlberg}, \enquote{\bibinfo{title}{Convolutional sparse
  representations with gradient penalties}},  \bibinfo{journal}{ICASSP, IEEE
  Int. Conf. Acoust., Speech and Sig. Proc. - Proceedings} \textbf{2018-},
  \bibinfo{pages}{8462151} (\bibinfo{year}{2018}).

\bibitem{bianco2018a}
\bibinfo{author}{M.~J. Bianco} and \bibinfo{author}{P.~Gerstoft},
  \enquote{\bibinfo{title}{Travel time tomography with adaptive dictionaries}},
   \bibinfo{journal}{IEEE Trans. Comput. Imaging} \textbf{4}(4),
  \bibinfo{pages}{499--511} (\bibinfo{year}{2018}).

\bibitem{grosse2007a}
\bibinfo{author}{R.~Grosse}, \bibinfo{author}{R.~Raina},
  \bibinfo{author}{H.~Kwong}, and \bibinfo{author}{A.~Y. Ng},
  \enquote{\bibinfo{title}{Shift-invariant sparse coding for audio
  classification}},  \bibinfo{journal}{Proc. Conf. on Uncert. in Art. Int.}
  \bibinfo{pages}{149--158} (\bibinfo{year}{2007}).

\bibitem{m2008a}
\bibinfo{author}{M.~M{\o}rup}, \bibinfo{author}{L.~K. Hansen},
  \bibinfo{author}{S.~M. Arnfred}, \bibinfo{author}{L.-H. Lim}, and
  \bibinfo{author}{K.~H. Madsen}, \enquote{\bibinfo{title}{Shift invariant
  multi-linear decomposition of neuroimaging data}},
  \bibinfo{journal}{Neuroimage} \textbf{42}(4), \bibinfo{pages}{1439--1450}
  (\bibinfo{year}{2008}).

\bibitem{batenkov2017a}
\bibinfo{author}{D.~Batenkov}, \bibinfo{author}{Y.~Romano}, and
  \bibinfo{author}{M.~Elad}, \enquote{\bibinfo{title}{On the global-local
  dichotomy in sparsity modeling}},  \bibinfo{journal}{Applied and Numerical
  Harmonic Analysis} \bibinfo{pages}{1--53} (\bibinfo{year}{2017}).

\bibitem{papyan2017b}
\bibinfo{author}{V.~Papyan}, \bibinfo{author}{Y.~Romano}, and
  \bibinfo{author}{M.~Elad}, \enquote{\bibinfo{title}{Convolutional neural
  networks analyzed via convolutional sparse coding}},  \bibinfo{journal}{J.
  Machine Learning Research} \textbf{18}, \bibinfo{pages}{1--52}
  (\bibinfo{year}{2017}).

\bibitem{cohen2018sparse}
\bibinfo{author}{R.~Cohen} and \bibinfo{author}{Y.~C. Eldar},
  \enquote{\bibinfo{title}{Sparse convolutional beamforming for ultrasound
  imaging}},  \bibinfo{journal}{IEEE Trans. Ultras, Ferroel., and Freq.
  Control} \textbf{65}(12), \bibinfo{pages}{2390--2406} (\bibinfo{year}{2018}).

\bibitem{llu2020a}
\bibinfo{author}{F.~Lluís}, \bibinfo{author}{P.~Martínez-Nuevo},
  \bibinfo{author}{M.~Bo~Møller}, and \bibinfo{author}{S.~Ewan~Shepstone},
  \enquote{\bibinfo{title}{Sound field reconstruction in rooms: Inpainting
  meets super-resolution}},  \bibinfo{journal}{J. Acoust. Soc. Am.}
  \textbf{148}(2), \bibinfo{pages}{649} (\bibinfo{year}{2020}).

\bibitem{grumiaux2021a}
\bibinfo{author}{P.-A. Grumiaux}, \bibinfo{author}{S.~Kitić},
  \bibinfo{author}{L.~Girin}, and \bibinfo{author}{A.~Guérin},
  \enquote{\bibinfo{title}{A survey of sound source localization with deep
  learning methods}},  \bibinfo{journal}{J. Acoust. Soc. Am.} \textbf{152}(1),
  \bibinfo{pages}{107--151} (\bibinfo{year}{2022}).

\bibitem{gerstoft2018a}
\bibinfo{author}{P.~Gerstoft}, \bibinfo{author}{C.~F. Mecklenbr\"auker},
  \bibinfo{author}{W.~Seong}, and \bibinfo{author}{M.~Bianco},
  \enquote{\bibinfo{title}{Introduction to compressive sensing in acoustics}},
  \bibinfo{journal}{J. Acoust. Soc. Am.} \textbf{143}(6),
  \bibinfo{pages}{3731--3736} (\bibinfo{year}{2018}).

\bibitem{candes2008a}
\bibinfo{author}{E.~J. Candes} and \bibinfo{author}{M.~B. Wakin},
  \enquote{\bibinfo{title}{An introduction to compressive sampling}},
  \bibinfo{journal}{IEEE Signal Process. Mag.} \textbf{25}(2),
  \bibinfo{pages}{21--30} (\bibinfo{year}{2008}).

\bibitem{boyd2010admm}
\bibinfo{author}{S.~Boyd}, \bibinfo{author}{N.~Parikh},
  \bibinfo{author}{E.~Chu}, \bibinfo{author}{B.~Peleato}, and
  \bibinfo{author}{J.~Eckstein}, \enquote{\bibinfo{title}{Distributed
  optimization and statistical learning via the alternating direction method of
  multipliers}},  \bibinfo{journal}{Foundations and Trends in Machine Learning}
  \textbf{3}(1), \bibinfo{pages}{1--122} (\bibinfo{year}{2010}).

\bibitem{heide-2015-fast}
\bibinfo{author}{F.~Heide}, \bibinfo{author}{W.~Heidrich}, and
  \bibinfo{author}{G.~Wetzstein}, \enquote{\bibinfo{title}{Fast and flexible
  convolutional sparse coding}}, in \emph{\bibinfo{booktitle}{Proc. IEEE Conf.
  on Comp. Vision and Pattern Rec. (CVPR)}} (\bibinfo{year}{2015}), pp.
  \bibinfo{pages}{5135--5143}.

\bibitem{wohlberg2016b}
\bibinfo{author}{B.~Wohlberg}, \enquote{\bibinfo{title}{Efficient algorithms
  for convolutional sparse representations}},  \bibinfo{journal}{IEEE Trans.
  Image Processing} \textbf{25}(1), \bibinfo{pages}{7308045}
  (\bibinfo{year}{2016}).

\bibitem{wohlberg2016a}
\bibinfo{author}{B.~Wohlberg}, \enquote{\bibinfo{title}{Boundary handling for
  convolutional sparse representations}},  \bibinfo{journal}{Proceedings -
  International Conference on Image Processing, Icip} \textbf{2016-},
  \bibinfo{pages}{7532675, 1833--1837} (\bibinfo{year}{2016}).

\bibitem{wohlberg2017a}
\bibinfo{author}{B.~Wohlberg} and \bibinfo{author}{P.~Rodriguez},
  \enquote{\bibinfo{title}{Convolutional sparse coding: Boundary handling
  revisited}},   (\bibinfo{year}{2017}).

\bibitem{hansen1998a}
\bibinfo{author}{P.~C. Hansen}, \emph{\bibinfo{title}{Rank-Deficient and
  Discrete Ill-Posed Problems: Numerical Aspects of Linear Inversion}}
  (\bibinfo{publisher}{SIAM}, \bibinfo{address}{Philadelphia},
  \bibinfo{year}{1998}), pp. \bibinfo{pages}{1--16}.

\bibitem{Note1}
\bibinfo{note}{See the code repository \dourl{https://github.com/manvhah/convolutional_plane_waves} to run experiments.}

\bibitem{heide2015a}
\bibinfo{author}{F.~Heide}, \bibinfo{author}{W.~Heidrich}, and
  \bibinfo{author}{G.~Wetzstein}, \enquote{\bibinfo{title}{Fast and flexible
  convolutional sparse coding}},  \bibinfo{journal}{Proc. IEEE Conf. on Comp.
  Vision and Pattern Rec. (CVPR)} \textbf{07-12-}, \bibinfo{pages}{7299149,
  5135--5143} (\bibinfo{year}{2015}).

\bibitem{wohlberg-2014-efficient}
\bibinfo{author}{B.~Wohlberg}, \enquote{\bibinfo{title}{Efficient convolutional
  sparse coding}}, in \emph{\bibinfo{booktitle}{Proc. IEEE Int. Conf. on
  Acoust., Speech, and Sig. Process. (ICASSP)}} (\bibinfo{year}{2014}), pp.
  \bibinfo{pages}{7173--7177}.

\bibitem{wohlberg-2017-sporco}
\bibinfo{author}{B.~Wohlberg}, \enquote{\bibinfo{title}{{SPORCO}: {A} {P}ython
  package for standard and convolutional sparse representations}}, in
  \emph{\bibinfo{booktitle}{Proceed. of the 15th Python in Science Conf.}},
  \bibinfo{address}{Austin, TX, USA} (\bibinfo{year}{2017}), pp.
  \bibinfo{pages}{1--8}.

\bibitem{hahmann2021b}
\bibinfo{author}{M.~Hahmann}, \bibinfo{author}{S.~A. Verburg}, and
  \bibinfo{author}{E.~Fernandez-Grande}, \enxquote{\bibinfo{title}{{Acoustic
  frequency responses in a conventional classroom}}} 
  Dataset \dodoi{10.11583/DTU.13315286} (\bibinfo{year}{2021}).

\end{thebibliography}

\ifdefined\appendcaptions
\section*{Figure captions}
\renewcommand{\labelenumi}{Figure \arabic{enumi}:}
\begin{enumerate}
\setlength{\itemindent}{5ex}
\item{\captionone}
\item{\captiontwo}
\item{\captionthree}
\item{\captionfour}
\item{\captionfive}
\item{\captionsix}
\item{\captionseven}
\item{\captioneight}
\end{enumerate}
\fi

\end{document}